\documentstyle[preprint,eqsecnum,aps]{revtex}

\begin{document}
\draft
\title{An exact diagonalization demonstration of incommensurability
and rigid band filling for N holes in the $t - J$ model}
\author{R.J. Gooding, K.J.E. Vos}
\address{Dept. of Physics, Queen's University,\\
Kingston, Ontario K7L 3N6}
\author{P.W. Leung}
\address{Dept. of Physics, Hong Kong University of Science and Technology,\\
Clear Water Bay, Hong Kong}
\date{\today}
\maketitle
\newpage
\begin{abstract}
We have calculated $S(\vec q)$ and the single particle
distribution function $<n_{\vec q}>$ for N holes in the
$t - J$ model on a non--square $\sqrt {8} \times \sqrt{32}$ 16--site lattice
with
periodic boundary conditions. We justify the
use of this lattice by appealing to results
obtained from the
conventional $4 \times 4$ 16--site cluster, and an undoped 32--site
system,
each having the full square symmetry of the bulk. This new cluster has a
high density of
$\vec k$ points along the diagonal of reciprocal space, viz.
along $\vec k = k (1,1)$.
The results clearly demonstrate that when the single
hole problem has a ground state with a system momentum
of $\vec k = ({\pi\over 2},{\pi\over 2}),$ the resulting ground state for
N holes involves a shift of the peak of the system's structure factor away
from the antiferromagnetic state $\vec q = (\pi,\pi)$. This shift effectively
increases
continuously with N.
When the single hole problem has
a ground state with a momentum that is not equal to $\vec k =
({\pi\over 2},{\pi\over 2}),$
something that may easily be accomplished through the use of
the $t - t^\prime - J$ model with $t^\prime/t$ small and positive,
then the above--mentioned incommensurability for N holes is not found -- the
maximum of $S(\vec q)$ remains at $\vec q = (\pi,\pi)$ for all N.
The results for the incommensurate ground states
can be understood in terms of rigid--band filling: the effective
occupation of the single hole $\vec k = (\pm{\pi\over 2},\pm{\pi\over 2})$
states is demonstrated by the evaluation of the single particle momentum
distribution
function $<n_{\vec q}>$. Unlike previous studies, we show that for the
many hole ground state the occupied momentum states are indeed
$\vec k = (\pm {\pi\over 2},\pm {\pi\over 2})$ states.
These conclusions are in agreement with the predictions for the spiral
phase made by Shraiman and Siggia. Further, our results demonstrate that
in some instances important results of moderately doped $CuO_2$ planes can be
predicted
from a knowledge of the properties of weakly doped planes.

\end{abstract}

\newpage
\section{Introduction:}
\label{sec:intro}

The $CuO_2$ plane based high--temperature superconductors
have anomalous normal state properties, and it is probable
that a complete theory of the superconducting instability
will first require a theory of this phase. One part
of the normal state puzzle involves the spin dynamics,
and in $La_{2-x}Sr_xCuO_4$ for $x=0.075,~0.14$ and $x=0.15$ recent
experiments\cite{cheong,mason,thurston} have demonstrated the existence of
incommensurate magnetic fluctuations. An explanation of these results
is an outstanding theoretical problem.

One theoretical starting point for these materials is the strong
coupling limit of the Hubbard model\cite{pwa}, and it has been
argued that the simpler $t - J$ model\cite{zhangrice} adequately
represents the important low--energy physics of these systems.
Then, the question is: do the predictions of the normal state
properties extracted from the $t - J$ model agree with experiment?
Here we shall focus on the predictions of the magnetic features
of these systems that can be made from
the strong--coupling limit.  Numerous theoretical
treatments\cite{spiral,zaanen,lee,tsanovic,schulz,inui}
of this problem have indeed suggested that some form of
a magnetic instability towards an incommensurate phase may arise in this
model. Unfortunately, not all of these theories agree with one
another, so more work is required to clarify the situation.

As elaborated in a review by Dagotto\cite{elbioreview},
one avenue by which theorists may scrutinize theoretical predictions
involves the use of exact diagonalization techniques. This allows
for the complete determination of all eigenstates of a given
system. The limitation of this method is simply the rather small
systems that can be studied, and thus comparisons of theory
to experiments on bulk systems can be quite limited. Progress is being made,
and
recent sophisticated techniques have been developed to treat somewhat
larger Hilbert spaces. One finite--size scaling study\cite{fss} of
a doped $t - J$ model yielded the encouraging result that the commonly used
$4~\times~4$ 16--site square lattice has only small finite size effects, at
least
for one hole.

An exact diagonalization study of the $t - J$ model for
a variety of carrier densities was performed by Moreo
{\it et~al.} \cite{moreo}. These studies were conducted before the
above--mentioned experiments\cite{cheong,mason,thurston},
and thus Moreo {\it et~al.} focussed on a search for an incommensurate
phase that was stable in the thermodynamic limit. We now
know that the incommensurability is only found in the
spin dynamics, and thus different questions
are important. To be specific: (i) what kind of incommensurability
(if any) is actually found in the $t - J$ model, and (ii) what are the
underlying states that the carriers occupy when in such a state?

To make clear the relevance of the latter question, consider that
using the $t - J$ model Shraiman and Siggia\cite{spiral} have
predicted the development of an incommensurate spiral phase
as a $CuO_2$ plane is doped
away from half filling. Implicit in the development of their
theory of the spiral phase is the assumption that at very low
doping levels the carriers approximately exist in momentum states
corresponding to the ground state of the one--hole problem.
(The one--hole problem has been studied using a variety of
techniques, and it is well known\cite{trugman,ss1,kaneleeread}
that the ground state has a system momentum $\vec k =
(\pm{\pi\over 2},\pm{\pi\over 2})$.
This has been confirmed by various exact diagonalization
studies\cite{elbioreview}, including the finite--size scaling work\cite{fss}
mentioned above.) Thus, the resilience of some form of rigid
band filling around so--called hole pockets is crucial if the
instability suggested by Shraiman and Siggia is to be validated.
Since recent photoemission work\cite{liu}
on $YBa_2Cu_3O_{6.35}$ have found partial evidence for the hole pocket
picture in the low doping regime, this question is clearly
very important.

The potential success of Ref. 6 in predicting the magnetic features
of the moderately doped high $T_c$ superconductors is related to
an even bigger and more important question: can lessons learned from
studying the properties of the weakly doped $CuO_2$ planes, $e.g.$
the single hole problem, be used to correctly extrapolate to higher doping
levels?
We now know that at least for small but nonzero doping levels this may be
the case: For one hole localized by a divalent $Sr$ impurity, theory
has predicted the ground state\cite{szczepanski,rabebhatt,skyrmion}. The
magnetic
component of the ground state was identified\cite{skyrmion},
and based on the above--mentioned semiclassical field theory of Shraiman
and Siggia\cite{ss1} it was realized that a two--fold degenerate non--coplanar
spin texture was present. (This spin state may be thought of as that
resulting from a particular superposition of ferromagnetic bonds\cite{aharony}
in
a 2D antiferromagnetic lattice\cite{skyrmion}.) Then, experiment
showed\cite{keimer}
that such a model\cite{polarons,nskyrmions} correctly reproduced the
zero--temperature antiferromagnetic correlation length. More recently
it was demonstrated that this is also true for nonzero
temperatures\cite{nskyrmions}.
Lastly, using $La$ NQR\cite{chou}
it has recently been observed that at very low
dopings and low temperatures, coexisting
with long--ranged antiferromagnetic order is a transverse
spin freezing - the temperature at which the freezing occurs may be
analytically predicted \cite{noha} using either the semiclassical
field theory, or {\it accurately} predicted
numerically \cite{noha} using the model employed in Ref. 25. These successes
suggest that
perhaps one might be able to continue to extrapolate the semiclassical theory
to even higher
doping levels, and this possibility, along with the experiments\cite{cheong}
consistent with the spiral phase\cite{spiral}, were the initial motivation for
this paper.

Thus, here we will present two correlation functions measured using
ground states obtained from an exact diagonalization study of the $t - J$ model
for metallic densities of carriers.
We focus on the two questions mentioned above, $viz.$ (i) is there
any evidence that at non--zero doping levels the ground state for the $t - J$
model displays any hints of incommensurability, and (ii) if so,
which single--particle states are occupied in the incommensurate phase.
Our results will provide some justification for the spiral phase
arguments made by Shraiman and Siggia, as well as for the similarities of the
ground states for the weak and moderate doping regimes.
Our paper is organized as follows. In \S II we introduce the cluster on
which the exact diagonalization determination of the ground state
was accomplished. We justify the use of this non--standard, non--square
lattice by
appealing to exact diagonalization results obtained on other lattices
possessing the full
square symmetry of the plane. In \S III we describe the magnetic properties of
N holes subject to the $t - J$ model for this cluster; we focus on the static
structure factor,
$S(\vec q)$, and show that as the cluster is doped, the peak in
$S(\vec q)$ shifts away from the antiferromagnetic wave vector.
Then, in \S IV we consider
the $t - t^\prime - J$ model, and demonstrate what happens when the single hole
ground state has a crystal momentum that is
not at $\vec k = (\pm {\pi \over 2}, \pm {\pi \over 2})$:
simply, the above mentioned incommensurability is no longer found.
In \S V we analyse the occupation of the
single particle momentum states -- we show that for the one, two, three, and
four hole systems,
the occupation of
the associated momentum states is not unlike the situation that would be
predicted via rigid band filling arguments. Finally, in \S VI we discuss our
numerical
results, focusing on comparisons to other theories and previous exact
diagonalization
studies.

\section{Description and Properties of Non--Square 16--Site Cluster:}
\label{sect:cluster}

Exact diagonalization is now a familiar technique by which studies
of systems with small Hilbert spaces can be carried out. For two--dimensional,
$S = {1\over 2}$ quantum spin systems (including doped
quantum antiferromagnets, a strong--coupling model
of the high--temperature superconductors) the most
commonly studied Hilbert space corresponds to a square $4 \times 4$
cluster with periodic boundary conditions.
Since a cluster of spins is supposed to represent a portion of the bulk
of the crystal, it has always been thought to be
desirable to ensure that the symmetry of
the bulk be maintained in the cluster. In this section we will introduce
a non--square cluster of 16 spins with periodic boundary
conditions. Then, we will justify the use of this lattice
by comparing to results for clusters which have the
full square symmetry, and, in fact, we will see that some of the unphysical
results obtained with the square 16--site cluster are eliminated when
our non--square cluster is used.

Figure 1 shows a cluster of sites which represent a small
portion of a bulk, two--dimensional square lattice; in total,
it contains 32 sites. Also, this cluster has the full $4mm$
point group symmetry of the bulk lattice (though obviously
not the translational periodicity of the infinite square
lattice). We impose periodic boundary conditions on this cluster,
and this yields the reciprocal lattice vectors shown in Fig. 2a.

Our non--square 16--site lattice is also shown in Fig. 1 -- it
is outlined by the rectangle elongated along the (1,1) direction,
and may be referred to as a $\sqrt{8} \times \sqrt{32}$ lattice.
Clearly, it has a lower point group symmetry, $viz.$ it only possesses
a centre of inversion symmetry. Imposing periodic
boundary conditions, the reciprocal lattice vectors for this
cluster are shown in Fig. 2b -- note that due to the lack of square
symmetry of this cluster  $\vec k = (k_{x},k_{y})$
is not necessarily equivalent to $(k_{y},k_{x})$.

Our motivation for choosing this cluster is two fold. Firstly,
we wish to dope this lattice and determine whether or not there
is any sign of incommensurability in the many--hole ground state.
If the ordering wave vector
shifts continuously (with doping) away from the ordering wave vector for the
commensurate antiferromagnetic insulator state,
$viz.$ $\vec q = (\pi, \pi)$,
then we should employ a cluster that has as many reciprocal lattice
vectors close to $(\pi, \pi)$ as possible. As seen in Fig. 2b, our
non--square 16--site lattice has a multitude of $k$ points along
the zone diagonal that are close to $(\pi, \pi)$,
$viz.$ $\vec k = (\pi, \pi), ({3\pi\over 4},{3\pi\over 4}), ({\pi\over
2},{\pi\over 2})$,
and $\vec k = ({\pi\over 4},{\pi\over 4})$.
Secondly, if any incommensurability is found in our studies, we
wish to understand the origin of the possible instability that leads
to the incommensurate state. Thus, if we are going to scrutinize the above
mentioned theories, we should not eliminate the proposed progenitors of the
incommensurability. Here we shall focus on whether or not the holes
tend to form many-hole wave functions that are
essentially constructed from a rigid band filling of the
associated one--hole ground states. The important (low energy) one
hole states are $\vec k = (\pi,0),$ and $\vec k = ({\pi\over 2},{\pi\over 2})$,
and as shown in Fig. 2b, our non-square 16--site cluster does indeed
possess both of these reciprocal lattice vectors. Thus, the 16--site
non--square
cluster shown in Fig. 1 is ideal for our purposes {\it if} its
lack of square symmetry does not produce any anomalous results;
we now show that this is indeed the case.

\subsection{Behaviour of the Undoped Non--Square 16--Site Cluster:}
\label{sect:undoped}

We have evaluated the ground state, and first excited state, for the
32 and 16--site clusters shown in Fig. 1, as well as for the
common $4 \times 4$ square cluster, for the antiferromagnetic Hamiltonian
\begin{equation}
H = J \sum_{<ij>} \vec S_i \cdot \vec S_j
\end{equation}
\label{equation:heisenberg}
when an $S = {1\over 2}$ spin is placed at every site of the cluster,
and periodic boundary conditions are used. For all three clusters
the ground state was a  $\vec k = 0$ singlet; for the two
16--site clusters, the ground state energies per spin were found to be
very close to one another: -.7018 for the $4 \times 4$ cluster,
and -.7085 for the non--standard 16--site cluster. Further, the
first excited state for all three clusters was always found to be
a $\vec k = (\pi, \pi)$ triplet; for the two
16--site clusters the mass gap (per site) was found
to be very close: .0723 for the $4 \times 4$ cluster, and .0740 for the
non--square cluster.

We are interested in the magnetic structure factor of the doped lattice; thus,
we must be sure that the non--square 16--site cluster does not yield
any anomalous results for this quantity.
The magnetic structure factor corresponds to
\begin{equation}
S (\vec q) = {1\over N}~< GS^n_{\vec k} |\Big[~\sum_{{\vec i, \vec j}}
{}~e^{i \vec q \cdot (\vec i - \vec j)} ~\vec S_{\vec i} \cdot
\vec S_{\vec j}~       \Big] | GS^n_{\vec k} >
\end{equation}
\label{equation:sofq}
where $| GS^n_{\vec k} >$ is the ground state for $n$ holes
having system momentum $\vec k$, and N is the total number of sites.
In Fig. 3 we show this quantity for all three clusters; since
the reciprocal lattice points do not always overlap, the comparison can
only be made at certain points.
It is clear that all three clusters give the same general features. Further,
and most importantly to this study, for the $\vec q$ along the zone diagonal,
the 32--site cluster and our non-square
16--site cluster have very similar static structure factors. For example, for
$\vec q = ({\pi \over 2}, {\pi\over 2})$ all three clusters have near
identical values of $S(\vec q)$.

It is apparent from these results that no anomalous features arise when the
non--square cluster
is used for an {\it undoped} Heisenberg Hamiltonian; we now consider
the doped cluster.

\subsection{Behaviour of the Doped Non--Square 16--Site Cluster:}
\label{sect:doped}

We have investigated the $t - J$ model, defined by
\begin{equation}
H = -t~\sum_{<i,j>\sigma} \Big( \tilde c_{i\sigma}^\dagger
\tilde c_{j\sigma} + h.c. \Big) + J \sum_{<ij>}
(\vec S_i \cdot \vec S_j - {1\over4} n_i n_j)
\end{equation}
\label{equation:tJ}
where we choose $t = 1$ and $J = .4$, as representative of
a $CuO_2$ plane. The operators $\tilde c_{i\sigma}^\dagger, \tilde c_{i\sigma}$
are
the creation and annihilation operators, respectively, corresponding to the
Hilbert space which has been reduced
by having had all doubly occupied sites integrated out; the
notation $<i,j>$ implies that only near--neighbour pairs are
summed over.

We added a single hole to the half--filled, antiferromagnetic insulator;
then, the minimum energy state was determined for every allowed system momentum
for both sixteen site clusters. The results, cast in the form of a
``band" structure, are shown
in Fig. 4. The variation of energy with respect to wave vector
is seen to be similar for the two clusters,
although the band width for the non--square cluster is smaller than for the
square cluster.

One intriguing {\it advantage} to the use of the non--square cluster is quickly
recognized from these results. To be specific, for one hole and the $t - J$
Hamiltonian,
use of the $4 \times 4$ square cluster yields the entirely unphysical
result that all states with system momenta $\vec k = (\pi,0)$ and $({\pi\over
2},
{\pi\over 2})$ (and, of course, those $\vec k$ points related to these by
the $4mm$ square symmetry) are degenerate; a proof of this fact
may be found elsewhere\cite{frenkel}. This is unfortunate since these two
states will
be non--degenerate in the bulk limit. Further, theory predicts that these
two states are the two lowest energy states assumed by a single hole.
Our non--square 16--site cluster is very
useful in that it contains all of these $\vec k$ points, and also has a
sufficiently
small Hilbert space such that the one hole states can be accessed, {\it but}
there
is no artificial (geometry--imposed) degeneracy between $\vec k  = (\pi,0)$ and
$\vec k = ({\pi\over 2}, {\pi\over 2})$. From Fig. 4 it is seen that the single
hole
ground state for the non--square cluster is $\vec k = ({\pi\over 2}, {\pi\over
2})$, and $(\pi, 0)$
is an excited state; this is consistent with the conclusions that have been
reached
regarding the single hole problem\cite{trugman,ss1}. (One persistent
disadvantage
found when using this cluster for one hole is that for $k_x - k_y = \pm \pi$,
the
minimum energy states are degenerate; the same phenomenon occurs
for the square 16--site cluster. Only the degeneracy along $k_x + k_y
= \pm \pi$ found in the square 16--site cluster is removed when we use the
non--square
cluster.)

Summarizing this section, we have introduced a non--square 16--site
cluster with periodic boundary conditions. Important reciprocal lattice
points are present in this lattice, and in comparison to the
square $4 \times 4$ 16--site cluster certain artificial degeneracies
are lifted. No anomalous results were found for the undoped or singly
doped non--square cluster.

\section{Incommensurability vs. Number of Holes:}
\label{sec:incommensurability}

We have used exact diagonalization to find the ground state of the $t - J$
Hamiltonian
for the non--square 16--site cluster for one through four holes\cite{leung};
this corresponds
to doping levels of $x = .0625$ to $x = .25$, and covers the experimental range
of
interest for systems that have displayed
incommensurability\cite{cheong,mason,thurston}.

The ground states, for $t = 1$ and $J = .4$, for $N =$ 1 and 2 holes have
crystal
momenta $\vec k = \pm ({\pi\over 2},{\pi\over 2}),$ and $\pm (\pi,\pm \pi),$
respectively.
For 3 holes the ground state is highly degenerate at the following reciprocal
lattice points: $\pm(-\pi,0), \pm(-{3\pi\over 4}, {\pi\over 4}), \pm(-{\pi\over
2}, {\pi\over 2}),
\pm(-{\pi\over 4}, {3\pi\over 4}),$ and $\pm(0,\pi)$. For 4 holes the ground
state
is found to correspond to momenta
$\pm ({\pi\over 2}, {\pi\over 2})$.

We have calculated the static structure factor, defined in
Eq. (2.2), and our results are shown in Fig. 5. The maximum of $S (\vec q)$
occurs at a wave vector which shifts
from $(\pi,\pi)$ (the antiferromagnetic wave vector) for one and two holes, to
$({3\pi\over 4},{3\pi\over 4})$ for three holes, to $({\pi\over 2},{\pi\over
2})$
for four holes. (Note that for ground states with non--zero
crystal momenta, one should perform an
average over the set of ground state wave functions that are degenerate (due to
the degeneracy of the ground state with respect to differing $\vec k$ points);
here, for 1, 2, and 4 holes, due to the lack of mirror symmetry
about the $x$ and $y$ axes for our non--square cluster, this does not change
the
results that are obtained when performing this average, $viz.$ only $\vec k$
and $-\vec k$ are degenerate, and $S (\vec q)$ is insensitive to which of these
ground state eigenfunctions is used. For 3 holes the same structure
factor is obtained for all of the degenerate
wave vectors.)

As the second hole is added, all that happens is a reduction of the
antiferromagnetic
correlations - this may also be seen in another correlation function,
$viz.$ the relative decrease of the near--neighbour spin--spin correlation
function $<\vec S_i \cdot \vec S_j >$.
Then, for three and four holes, an essentially continuous shift
in peak position occurs; the continuous shift in wave vector is seen to mimic
the
experiments of Cheong $et~al.$\cite{cheong}. For more than four holes, $S (\vec
q)$
is essentially flat, indicating the effective loss of magnetic correlations
in the heavily doped materials.

It would be desirable to be able to perform the same search for
incommensurability on
a lattice with a high density of $\vec k$ points around $(\pi,\pi)$ such that
the neighbouring $\vec k$ points were along the $(1,0)$ and/or $(0,1)$
direction;
this is the direction of the incommensurate shifts found
experimentally\cite{cheong}.
However, the only lattice (with a small number of sites, and thus appropriate
for
exact diagonalization studies of a multiply doped cluster) is that of a ladder
of width two - this
cluster would in no way approximate the bulk lattice, and thus we must be
content
with a search for incommensurabilities along the zone diagonal. Further, it has
been suggested, in a weak coupling theory, that one cannot reproduce the
experimentally
observed shifts in a one--band model; instead, a three-band model
is required to produce the necessary nesting\cite{littlewood}. Even if we had
used
a three--band model, the nature of our cluster still restricts us to the set of
$\vec k$ points
explored here, and thus we do not believe studies of $S (\vec q)$ on finite
clusters
in the strong coupling limit could yield more information on the
incommensurability
than we have found {\it until} the technical obstacles associated with doping a
32, or 36--site
cluster\cite{fss,ziman} with many holes are overcome - this may {\it never} be
possible.
Further, only with such progress could the finite--size scaling be carried
out to scrutinize the observation\cite{mason} that a very weak logarithmic
maximum
of $S(\vec q)$ exists at the incommensurate wave vectors.

\section{Incommensurability in the $\lowercase{t} -\lowercase{t}^\prime - J$
model:}
\label{sec:tprime}

We have considered the ground state of the $t - t^\prime - J$ model.
This model corresponds to the Hamiltonian of Eq. (2.3) augmented
with a next nearest neighbour hopping:
\begin{equation}
H_{t^\prime} = -t^\prime~\sum_{<i,i^\prime>\sigma}
\Big( \tilde c_{i\sigma}^\dagger
\tilde c_{i^\prime\sigma} + h.c. \Big)
\end{equation}
\label{tprimeH}
where $i^\prime$ is a next near neighbour to $i$. The inclusion
of this new term has been motivated in a variety of ways
\cite{hybertsen,goodingelser,shastry};
here it is extremely useful in showing
how the incommensurability demonstrated in the above section
is changed when the crystal momentum associated with the single
hole problem is shifted away from $\vec k = \pm ({\pi\over 2},{\pi\over 2})$.
For $t = 1.$ and $J = .4$, we have added a small positive $t^\prime$,
viz. $t^\prime = .2$, and found the ground state for our
non--square 16--site cluster. The ground state momentum is no
longer at $\vec k = \pm ({\pi\over 2}, {\pi\over 2})$, but now
is found to be at $\vec k = \pm (\pi, 0)$. This behaviour is consistent
with the band--structure predictions for a hole moving in an inert
background.

In Fig. 6 we show $S(\vec q)$ for 1, 2, 3, and 4 holes, for the
$t - t^\prime - J$ model, using the model parameters given above.
It is clearly seen that the maximum of the magnetic structure
factor is always at the antiferromagnetic wave vector, viz.
$\vec q = (\pi, \pi)$. As the cluster is progressively doped, all that
happens is a suppression of the antiferromagnetic correlations -
no shift of $S(\vec q)$ to wave vectors neighbouring the antiferromagnetic
$(\pi, \pi)$ is found. (As mentioned above, this is also found when
studying the near--neighbour spin--spin correlation function $< \vec S_i
\cdot \vec S_j >$.) This is in marked contrast to the behaviour
found in the above section: cf. Fig. 5. This simple demonstration seems
to suggest that in the strong--coupling limit the formation of an
incommensurate phase requires the single hole problem to have
its ground state momentum equal to $\vec k = ({\pi\over 2},{\pi\over 2})$.
We now examine the single particle momentum distribution function to show why
this is so.

\section{Momentum Distribution Functions:}
\label{sec:nofk}
In the previous two sections we have displayed results obtained
from exact diagonalization studies that provide evidence
for incommensurate correlations in the strong coupling
limit of a two--dimensional doped antiferromagnetic insulator {\it when}
the single hole ground state was located at $\vec k = \pm ({\pi\over 2},
{\pi\over 2})$. The question that naturally arises is: why
does the one hole state so profoundly affect the many hole features?
In this section we wish to show that one can also use
the exact diagonalization results to suggest the progenitor
of this incommensurability, and subsequently answer this question
via a study of the electron and hole momentum distribution functions.

Our approach is very similar to one employed
by Stephan and Horsch~\cite {stephanhorsch}, as well as that
more recently given in a very clear presentation made by
Ding~\cite {ding} - in our work we shall follow the notation
of Ding. One defines the electron distribution function by
\begin{equation}
< n_{\sigma} (\vec q) > =  < \tilde c_{\vec q \sigma}^\dagger  \tilde c_{\vec q
\sigma} >.
\end{equation}
Similarly, a hole momentum distribution function
can be defined:
\begin{equation}
< {\overline n}_{\sigma} (\vec q) > =  < \tilde c_{\vec q \sigma}  \tilde
c_{\vec q \sigma}^\dagger >.
\end{equation}
Note that in using this definition, the hole distribution function includes
a spin idex, a feature induced by the constraint of no
double occupancy - this property is explained by Ding\cite {ding}.
We wish to track the electron and hole occupations
as our cluster is doped from one to four holes. To be specific,
we wish to ascertain which electron and hole states are occupied as the
incommensurability found in the previous section develops.

If one examines these distribution functions for the $t^\prime = 0$ ground
states
discussed above, one must overcome more unphysical degeneracies; $e.g.$,
the four hole $< n_{\sigma} (\vec q) >$ has degenerate values
for $q_x - q_y = \pm \pi$.  Further, the analysis is greatly complicated
by the degeneracy (with respect to the wave vector) of the many--hole ground
states;
$e.g.$, the unphysical degeneracy of the 3 hole state.  This problem for the
non--square
16--site cluster is unique to the pure $t - J$ model. To remove it one can add
the second
near--neighbour hopping $t^\prime$ introduced in \S \ref {sec:tprime} - it
is known that this hopping amplitude is of opposite sign to that of the near
neighbour
hopping \cite{hybertsen}.
We have chosen $t^\prime / t = -.1$ for a number of reasons: (i) with this
addition
the degeneracies of the many hole ground states are lifted, (ii) the single
hole ground
state remains at $\vec k = \pm (\vec {\pi\over 2}, {\pi\over 2})$, and in
comparison
to the $t^\prime = 0$ system, the ordering of the low energy excited states is
not changed,
and (iii) the two and four hole ground states become $\vec k = 0$ states, a
property that
one certainly would expect a bulk system with an even number
of holes to possess.  As an example of the usefulness
of including the second near neighbour hopping, note
that when $t^\prime =0$, the three hole ground state on our
non--square 16--site cluster is degenerate at the following
wave vectors: $\pm (-\pi, 0), \pm (-{3\pi\over 4}, {\pi\over 4}),
\pm (-{\pi\over 2}, {\pi\over 2}), \pm (-{\pi\over 4}, {3\pi\over 4}),
\pm (0, \pi)$. Then, when $t^\prime = -.1$ is added,
one finds that this unphysical degeneracy is lifted and the ground state occurs
at
$\pm ({\pi\over 2}, -{\pi\over 2})$. We wish to stress that identical
conclusions to the
ones presented below can be reached for any small and negative $t^\prime$
\cite{lossofinc}.

For 1, 2, 3, and 4 holes in the $t - t^\prime - J$ model
with $t = 1$, $J = .4$, and $t^\prime = -.1$, on
our non--square 16--site cluster the ground state is found to
occur at $\vec k = \pm ({\pi\over 2}, {\pi\over 2}), (0,0),
\pm ({\pi\over 2}, - {\pi\over 2}),$ and $(0,0)$, respectively. Then, the
electron
and momentum distribution functions are as shown in Figs. 7, 8, 9, and 10.
For an odd number of holes we show both spin components
of the distribution functions; for an even number we show just one
of the two spin components for both the electrons and holes (since the
up and down spin distributions are equivalent). Various sum rules, etc.,
associated
with these numbers are discussed at length by Ding\cite {ding}.
The positioning of these numbers in the figures corresponds to the allowed
wave vectors of the non--square 16--site cluster, as shown in
Fig. 2b.

The one hole ground state (which was found in the subspace of the
total magnetization being ${1\over 2} \hat z$) clearly shows
the large occupation of the electron states within the antiferromagnetic
Brillouin
zone (defined by $|q_x| + |q_y| = \pi$) except for
electrons and holes at the wave vector of the
ground state, i.e.  $\vec k = \pm ({\pi\over 2}, {\pi\over 2})$;
only one of these electron states is found to be occupied.
This is the same result as was found by Ding\cite {ding}, and shows that
one may associate the momentum of the ground state, and the
momentum of the hole state, to be one and the same
{\it for~one~hole} even in the strong coupling limit.
Also, a comparison
of our Fig. 7 and Fig. 4 of Ding\cite {ding} provides evidence that our
non--square 16--site cluster has a momentum distribution
function that behaves in a similar fashion to that
found for the square $4 \times 4$ cluster.

The two hole ground state is a $\vec k = 0$
state (which was found in the subspace of zero
total magnetization), and as Fig.~8 shows,
the occupied electron states are within the antiferromagnetic Brillouin
zone except for states at the wave vector of the
one hole ground states, i.e.  $\vec k = \pm ({\pi\over 2}, {\pi\over 2})$.
Now, unlike the one hole ground state, only holes occupy these states.
This is precisely the distribution function that one would expect
based on rigid band filling arguments: the minimum energy states
for one hole are at $\vec k = \pm ({\pi\over 2}, {\pi\over 2})$, and
now for two holes both of these states are occupied by holes.
Also, this is a very different conclusion from
that reached by Stephan and Horsch~\cite {stephanhorsch} whose
results on a twenty site lattice suggested that the single hole
problem had little to do with the many hole ground state.
This may be understood in part because for their lattice the
important $\pm ({\pi\over 2}, \pm {\pi\over 2})$ states are not
present - see the discussion in \S \ref{sec:discussion}.

We have used a small negative $t^\prime$ to lift
the unphysical one hole degeneracies, and it is these degeneracies that
caused the difference between our results, shown in Fig. 8, and those of
Ding\cite{ding},
for two holes; this is a further example of the usefulness of including
$t^\prime$.
Ding found that the $t - J$ two hole ground state was degenerate
at $\vec k = (0,0), \pm (\pi,0),$ and $\pm (0,\pi)$ for the square $4 \times 4$
lattice. The inclusion of a small negative $t^\prime$ lifts
this degeneracy and makes the ground state a $\vec k = 0$ state. Then,
an analysis of the electron and hole distribution function clearly
shows the occupation of the $(\pm {\pi\over 2}, \pm {\pi\over 2})$ states,
consistent with rigid band filling.  This is to be compared with the occupation
of the
$\pm (\pi,0), \pm (0,\pi)$ states that Ding found in his $\vec k = 0$,
$t^\prime = 0$ ground state.

The three hole ground state (which was again found in the subspace of the
total magnetization being ${1\over 2} \hat z$) is at $\vec k
= \pm ({\pi\over 2}, - {\pi\over 2})$, and the electron and
hole distribution functions are shown in Fig. 9. In comparison to the
two hole case, we now see that the third hole occupies the same momentum
state as the crystal momentum of the ground state, while the first two
holes are still found to occupy the $\pm ({\pi\over 2}, {\pi\over 2})$ states.
This is again consistent with rigid band filling. To display this we
have provided the minimum energy states for one hole in the
$t - t^\prime - J$ model in Table I. Note that the first excited
state {\it within} the antiferromagnetic Brillouin zone is at $\vec k = \pm
({\pi\over 2},
-{\pi\over 2})$, and is thus
the state that one would expect the third hole to occupy. Figure 9 is
a vivid demonstration of the {\it hole~pockets} that one would expect
from rigid band filling arguments.

The four hole ground state (which was found in the subspace of
zero magnetization) is a $\vec k = 0$ state. It has
electron and momentum distribution functions, as displayed in Fig. 10,
very similar in character to those of the fewer hole states. Hole pockets
around the four
momenta $\pm ({\pi\over 2}, \pm{\pi\over 2})$ are clearly in evidence;
some small tendency towards an expansion of the pockets to form a closed fermi
surface
may be seen.
This is consistent with the assumption \cite {ss1} that
the band structure around the antiferromagnetic zone faces is very flat along
the zone boundary, but steep towards the $\vec k = 0$ point. These
results are also suggestive of a crossover from hole pocket states to
a Luttinger liquid\cite{lut}, although this simple set of data from a finite
cluster can in no way be considered to conclusively answer such an
important question.

\section{Discussion:}
\label{sec:discussion}

We have suggested the use of a non--square 16--site cluster which includes
{\it all} the important reciprocal lattice points for the one hole problem,
and lifts certain unphysical degeneracies.
We have shown that no anomalous results are found for this cluster, and
have doped it with a small number of holes. The magnetic structure factor
clearly shows the movement of its peak with carrier density reminiscent of an
incommensurate
phase. Since we are only working with a finite cluster, and are incapable
of doing a finite scaling analysis with these results, we cannot
be sure whether these correlations survive in the bulk limit, but experiments
suggest that only dynamical (i.e. short--ranged) correlations remain.
We will present the dynamic structure factor for this model in a future
publication, and this will allow for a more direct comparison with
experiment.

We have studied the electron and hole distribution functions for the
many hole problem. They provide clear evidence of the development of
hole pockets near the ground state wave vectors of the one hole problem;
as the doping increases it seems quite possible that the hole
pockets disappear, and a Fermi surface obeying Luttinger's theorem
results.
On the basis of an assumption of (i) the strong coupling limit,
(ii) the one hole ground state's character, $viz.$ that it is a
$\vec k = \pm ({\pi\over 2}, \pm{\pi\over 2})$ state producing long--ranged
dipolar spin distortions, and (iii) the existence
of such hole pockets, Shraiman and Siggia~\cite {spiral} proposed
the presence of an incommensurate spiral phase - our results strongly
support their theory.

Our results contrast with earlier studies of these same questions.
Firstly, Moreo $et~al.$ \cite{moreo} did not find robust evidence
of incommensurability when the doping level of a square $4 \times 4$
cluster described by the $t - J$ model. Our Fig. 5 seem to be very
direct evidence of such an underlying instability. In contrast to
our use of $S(\vec q)$, Dagotto $et~al.$ \cite{dagottoN} has used the dynamic
structure
factor for a variety of hole fillings, and did not find any evidence
of incommensurability. Clearly, our use of a cluster that includes all of
the important reciprocal lattice vectors, and a high density of $\vec k$
points around the antiferromagnetic  wave vector, has allowed us to
make a more direct study of this problem.

The work of Stephan and Horsch \cite{stephanhorsch} has been considered
by some \cite{horschstephan} to have clearly demonstrated that the single
hole problem has nothing to do with the higher doping levels of interest.
To be specific, their two hole work showed a Luttinger Liquid with
a clear Fermi surface, and no hint of hole pockets. Their work was
conducted on a number of different clusters. Our work brings into question
the absoluteness of these conclusions - we have clear evidence of
hole pockets, and a knowledge of the single hole ground and excited states are
found
to be all that is necessary to predict the behaviour of the single particle
momentum
distribution functions for many holes. Thus, the question that must be
answered is: how can two studies using the same technique (exact
diagonalization)
produce such totally different conclusions? We feel that because our
cluster has the important $\vec k = \pm ({\pi\over 2}, \pm{\pi\over 2})$
momentum states, and it is these states that are required to properly
incorporate the dipole--dipole interactions associated with the spiral
instability of Shraiman and Siggia \cite{spiral}, {\it and}, we do indeed
find an incommensurability in these ground states, consistent with
experiment, differing clusters lead to different hole--hole interactions, and
these
interactions {\it must} strongly depend on the momentum states that the holes
occupy.

To emphasize this latter point, we note that the work of Ding \cite {ding}
led him to conclude that some form of rigid band filling did indeed occur
for two holes. He, however, thought that the two single particle states that
combined
to produce the two hole ground state were $\vec k = \pm (\pi,0), \pm (0, \pi)$
states - no hole pockets are then produced.
We found that when the unphysical degeneracy of these
reciprocal lattice points and those at the faces of the antiferromagnetic
Brillouin zone are lifted (using a small negative $t^\prime$),
and thus a different form of rigid band filling,
one displaying hole pockets, is produced. This is again an example
of the strong dependence of the hole--hole interactions on the underlying
single
hole ground and first few excited states,
and the subsequent character of the many hole ground states.

Our results are clearly in support of some form of rigid band filling ($e.g.$,
see
Fig. 10), and thus suggest that knowledge gained from the study of the simpler
one hole problem
can (sometimes) be used to understand instabilities occurring at higher carrier
densities;
$e.g.$, the incommensurate spiral phase \cite {spiral}. This is similar to
conclusions
reached previously by one of us for the very weakly
doped insulator \cite{keimer,polarons,nskyrmions,chou,noha}, and lends credence
to studies of other aspects of this problem, $e.g.$ transport in the normal
state, that were based on an assumption of rigid band filling \cite{trugprl}
having begun with a strong--coupling description of doped $CuO_2$ planes.

\bigskip
\bigskip
\bigskip
\centerline{\bf {ACKNOWLEDGEMENTS:}}

We wish to thank T. Barnes, T. Mason, S. Trugman, and especially A.M.S.
Tremblay for helpful
comments. This work was supported by the NSERC of Canada.

\begin{figure}
\caption{The 32--site cluster; the rectangle outlines the non--square 16--site
cluster that we focus on in this paper; this cluster is seen to be half of the
32--site cluster.}
\label{clusterfig}
\end{figure}

\begin{figure}
\caption{Reciprocal lattice vectors for (a) the 32--site cluster,
and (b) our non--square 16--site cluster.}
\label{reciprocalspacefig}
\end{figure}

\begin{figure}
\caption{A comparison of the magnetic structure factor for the undoped
square $4 \times 4$ 16--site, our non--square 16--site, and the square
32--site, clusters.
The reciprocal lattice points are as follows: $\Gamma = (0,0),~ {\rm
X}=(\pi,0),$
and M=$(\pi,\pi)$.}
\label{comparingsofqfig}
\end{figure}

\begin{figure}
\caption{Band structures for one hole in the $4 \times 4$ and non--square
16--site clusters; we have used $t = 1$ and $J = .4$.}
\label{bandstructurefig}
\end{figure}

\begin{figure}
\caption{Magnetic structure factors for one, two, three, and four holes
for the $t - J$ model on the non--square 16--site cluster.}
\label{sofqvsNfig}
\end{figure}

\begin{figure}
\caption{Magnetic structure factors for one, two, three, and four holes
for the $t - t^\prime - J$ model on the non--square 16--site cluster, with
$t^\prime/t = .2$.}
\label{sofqtprimefig}
\end{figure}


\begin{figure}
\caption{Distribution functions for (a) electrons, and (b) holes, for the
single
hole problem. The ground state is degenerate at $\vec k = \pm ({\pi\over 2},
{\pi\over 2})$,
and here we show the distribution functions for the $k = (-{\pi\over 2},
-{\pi\over 2})$ state.
The upper (lower) numbers represent the spin up (down) components. The energy
parameters
of the $t - t^\prime - J$ model are $t = 1, t^\prime = -.1$, and $J = .4$. The
square
outlines the antiferromagnetic Brillouin zone.}
\label{oneholenofk}
\end{figure}

\begin{figure}
\caption{Distribution functions for electrons and holes, for the two
hole problem. The ground state is a $\vec k = 0$ state. The upper (lower)
numbers
represent the electrons (holes).}
\label{twoholenofk}
\end{figure}

\begin{figure}
\caption{Distribution functions for (a) electrons, and (b) holes, for the three
hole problem. The ground state is degenerate at $\vec k = \pm ({\pi\over 2},
-{\pi\over 2})$,
and here we show the distribution functions for the $k = ({\pi\over 2},
-{\pi\over 2})$ state.
The upper (lower) numbers represent the spin up (down) components.}
\label{threeholenofk}
\end{figure}

\begin{figure}
\caption{Distribution functions for electrons and holes, for the four
hole problem. The ground state is a $\vec k = 0$ state. The upper (lower)
numbers
represent the electrons (holes).}
\label{fourholenofk}
\end{figure}


\begin{references}
\bibitem{lossofinc} One caveat to choosing this value of $t^\prime$ is that the
incommensurability
found in the magnetic structure factor
is lost in that now the 3 hole ground state peaks at the antiferromagnetic
wave vector. However, at $\vec q = ({3\pi\over 4}, {3\pi\over 4})$ the magnetic
structure factor is only 1\% less than it is at $(\pi, \pi)$; we do not
consider this to have compromised our results. In fact, one can use a smaller
$| t^\prime |$, lift the unphysical 3 hole degeneracy, maintain the
incommensurability found in $S(\vec q)$, but the 2 and 4 hole
ground states are not $\vec k = 0$ states. Thus, we have used a value
of the second near neighbour hopping that produces this latter result. Using
a smaller $| t^\prime |$ does not change our results for the momentum
distribution
functions.
\end{references}
\end{document}